\documentclass[aip,jap, amsmath,amssymb,twocolumn]{revtex4}
\usepackage{graphics}
\usepackage{graphicx}
\usepackage{amsmath}
\usepackage{epstopdf}
\def\bnabla{\mbox{\boldmath $\nabla$}}

\begin{document}

\title{2D electrons floating on a suspended atomically thin dielectric}
\author{F. T. Vasko}
\email{ftvasko@gmail.com}
\affiliation{QK Applications, Monterey, CA 93940 }
\date{\today}

\begin{abstract}
The 2D electrons trapped in vacuum near the atomically thin dielectric (ATD, mono- or $N$-layer film of $h$-BN or transition metal dichalcogenide) are considered. ATD is suspended above the back gate and forms the capacitor which is controlled by the biased voltage determining 2D concentration, $n_{2D}$. It is found that the leakage current through ATD is negligible and effect of the polarizability of ATD is weak if $N\leq 5$. At temperatures $T=0.1\div$15 K and $n_{2D}=5\times 10^8\div 10^{10}$ cm$^{-2}$, one deals with the Boltzmann liquid of the macroscopic thickness $\sim$100 A. Due to bending of ATD the quadratic dispersion law of the flexural vibrations is transformed into the linear one at small wave vectors. The scattering processes of the electrons caused by these phonons or the monolayer islands on ATD are examined and the momentum and energy relaxation rates are analyzed based on the corresponding balance equations. The momentum relaxation times varies over orders of magnitude in the above region ($T$, $n_{2D}$) and $N$. The response may changed from the polaron transport, for a perfect single-layer ATD at low $T$ and high $n_{2D}$, to the high-mobility ($\geq 10^7$ cm$^2$/Vs) regime at high $T$ and low $n_{2D}$. The quasi-elastic energy relaxation due to the phonon-induced scattering is considered and the conditions for heating of electrons by a weak in-plane electric field are found.   
\end{abstract}
\maketitle

\section{Introduction }
Over the past 50 years, the transport and optical properties of two-dimensional electrons have been extensively studied in the metal-oxide-semiconductor \cite{1} and semiconductor \cite{2,3} heterostructures. Confinement of electrons in the layer of thickness $\gg a_B$ (the Borh radius) is  provided both by heterojunctions and electric fields applied through the metal gates. Because of a weakness of scattering for the electronic states with a macroscopic ($\geq$100 A) thickness, a lot of devices using 2D electrons are widely applied in electronics and optical communications. Beside of this, the 2D layer of electrons floating in vacuum on liquid helium have been demonstrated and investigated at low temperatures. \cite{4,5} For this system, confinement of electrons is due to the image potential and the electronic state of macroscopic ($\gg a_B$) thickness appears due to the weak polarisibility of He or due to a back gate under the He film. This is an effectively tunable system which show both a nearly ideal 2D transport, with the mobility exceeded the data for any solid-state device (see \cite{6} and references therein),  and the Wigner crystallization regime. \cite{7} In spite of this, a possible applications of the 2D electrons on liquid He, including a realization the qubit of a quantum computer suggested in Ref. 8, are restricted by a high sensitivity of the liquid substrate to an external perturbations. But it is not possible to replace of the He substrate by a bulk dielectric with  a  permittivity $\gg 1$, which leads to a localization of electrons at atomic distances $\sim a_B$ by the strong image force, \cite{9} and due to a surface  imperfection of  this substrate.
\begin{figure}[t]
\vspace{0.25 cm}
\begin{center}
\includegraphics[scale=0.21]{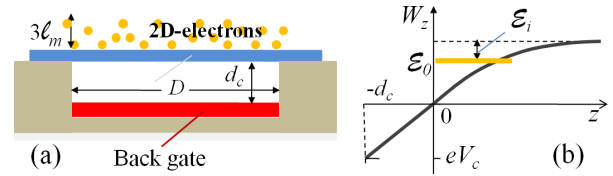}
\end{center}  %
\addvspace{-0.5 cm}
\caption{ (a) Sketch of the vacuum-isolated 2D electrons (orange balls formed layer of thickness $\sim 3\ell_m$) floated on an ATD  (blue, placed at $z=0$) suspended over trench of cross-section $D\times d_c$ under the bias voltage $V_c$ which is applied through the back gate (red, placed at $z=-d_c$). (b) Potential energy $W_z$ of the device with the ground level energy ${\cal E}_0$ and the energy of ionization ${\cal E}_i$.  }
\end{figure} 

During last decade an atomically thin dielectric (ATD) films, such as mono- and a few-layer $h$-BN or transition-metal dichalcogenides, have been extensively studied, see \cite{10,11} and references therein. Particularly, the electronic and heat transport in ATD, see \cite{12} and \cite{13}, as well as the mechanical and optical properties of ATD, see \cite{14} and \cite{15}, have been examined. A suspended mono- and a few-layer ATD have been studied \cite{16}, including the cases of large-size ATD placed onto a trench in the substrate. These results allow us to propose {\it a new possibility of the implementation of a 2D electron layer floating in vacuum over a suspended ATD.}

For a slow electron approaching to the atomically smooth ATD plane, the latter may be considered as an abrupt barrier in the transverse direction.  Because of the small thickness of the ATD, the image forces induced by electron are also negligible and it is possible to ensure the localization of 2D electrons in vacuum near the barrier using a back gate placed under the suspended ATD, see Fig. 1a. The mechanism of transverse localization suggested is more convenient then the case of 2D electrons  floating on He due to the replacement of the liquid substrate by the ATD and this device should be more stable and controllable. This paper addresses the questions on the conditions for localization of electrons at distances $\gg a_B$  above ATD, on the parameters of their energy spectrum, on the mechanisms of interaction with vibrations and roughness of ATD, and on the characteristics of the in-plane transport of 2D electrons. 

The consideration below involves the self-consistent calculations of the energy spectrum with the zero boundary condition at ATD, $z=0$ see the energy band diagram in Fig. 1b, and the estimates of the leakage current through ATD as well as its polarizability due to 2D electrons. Description of the flexural vibrations of ATD is performed within the elasticity theory \cite{17,18,19} taking into account a bending of suspended ATD. The transition probabilities  of the 2D electrons interacted with the flexural phonons or monolayer islands on ATD are found within the second-order perturbation theory. The in-plane transport is examined based on the balance equations for losses of the  drift velocity and the energy of 2D electrons. \cite{20} The dependencies of the momentum and energy relaxation times versus temperature, $T$, and concentration of 2D-electrons, $n_{2D}$, are analyzed for $T=0.1\div$15 K and $n_{2D}=5\times 10^8\div 10^{10}$ cm$^{-2}$. A nonlinear response on an in-plane electric field is also governed by the balance between the Joule  heating and the energy losses.

One can summarize the results obtained in the following points. {\bf (a)} The leakage current of the 2D electrons caused by the tunneling through ATD is not essential and the polarizability of ATD is negligible, so that the model of a narrow non-transparent barrier can be used. {\bf (b)} Flexural vibrations of suspended ATD are described by the linear dispersion law for the long wavelengths and by the quadratic one law for the short wave lengths with the crossing region determined by the bending of ATD. {\bf (c)} For the $(T, n_{2D})$-region considered (see above), one obtains the nondegenerate 2D electrons with a strong Coulomb interaction (the Boltzmann liquid floating on ATD). {\bf (d)} For an ATD without roughness and with low $n_{2D}$ at high $T$, the high-mobility transport takes place while for high $n_{2D}$ at low $T$ the phonon renormalization of mass is essential, i.e. one deal with the non-ideal plasma of polarons. {\bf (e)} The electron-phonon interaction is suppressed effectively in $N$-layer ATD and a roughness-induced scattering, with different dependencies on $n_{2D}$ and $T$, becomes essential. {\bf (f)} A nonlinear regime of in-plane transport due to the Joule heating appears already in weak field ($\sim$mV/cm) if the momentum relaxation via roughness and the ionization of 2D electrons are  negligible.

The paper is organized in the following way. In Sec. II we show that the leakage current through ATD and the effect of ATD's polarization on the 2D electron energy are negligible. The flexural vibrations of the ATD suspended a over long trench are examined in Sec. III. The self-consistent spectrum of electrons and their mechanisms of relaxation via the flexural phonons and via the roughness of ATD are described in Sec. IV.  In Sec. V we analyze the in-plane transport including the momentum and energy relaxation times and the nonlinear regime of response. The concluding remarks, the list of assumptions, and the discussion of current experimental context are given in the last section. 

\section{Suspended ATD }
Formation of a vacuum-insulated layer of 2D electrons above the ATD is possible under the two key conditions: {\it a)} a negligible leakage current between the 2D electrons and the back gate and {\it b)} a weak image force due to polarization of ATD. Here we address these conditions and demonstrate that a mono- or few-layer ATD is suitable for realization of the device suggested.

\subsection{Leakage rate }
First, we estimate the leakage rate caused by the tunneling of the 2D electrons into the quasi-3D states at $-Nl_0>z>-d_c$ through an ideal (without holes or capture centers) $N$-layer ATD; $l_0$ is the single-layer thickness and $d_c$ is the distance between ATD and back gate. Following \cite{20} (see Append. H) and \cite{21}, we introduce the tunneling matrix element $T_{0{\bf p},r{\bf p'}}= \left\langle {0{\bf p}} \right|\widetilde W_{\bf r} \left|{r{\bf p'}}\right\rangle$, where $\widetilde W_{\bf r}$ is the microscopic barrier potential in the region $-Nl_0 <z<0$ and $|0{\bf p}\rangle$ or $|r{\bf p'}\rangle$ are the under-barrier tails of wave functions at $z<0$ and $z>-Nl_0$, respectively. Here $r$ labels the quasi-discrete states in the region below ATD. Within the weak tunneling regime, the distribution of 2D-electrons is governed by the kinetic equation
\begin{equation}
\frac{df_{0{\bf p}t}}{dt}\! =\!\frac{2\pi}{\hbar}\!\sum\limits_{r{\bf p'}} {\left| {T_{0{\bf p},r{\bf p'}} } \right|}^2 \delta (\varepsilon_{0p}\! -\!\varepsilon_{rp'})\left( f_{0{\bf p}t}\! -\! f_{r{\bf p'}t} \right)
\end{equation}
with the initial condition $f_{0{\bf p}t\to 0}\to\widetilde f_\varepsilon$. Similar equation with the zero initial condition takes place for $f_{r{\bf p}}$. For the early stages of decay, when $f_{r{\bf p'}t}\ll f_{0{\bf p}t}$, temporal evolution of 2D electrons is described by
\begin{equation}
f_{0{\bf p}t}\! =\! \widetilde f_\varepsilon e^{-\Gamma_{\bf p}t} , ~~ \Gamma_{\bf p}\! =\!\frac{2\pi}{\hbar}\!\sum\limits_{r{\bf p}'}\!{\left| T_{0{\bf p},r{\bf p}'}\right|}^2 \delta (\varepsilon_{0p}\! -\!\varepsilon_{rp'} ) ,
\end{equation}
where $\Gamma_{\bf p}$ is the leakage rate described the exponential damping of 2D population.

Below we estimate $T_{0{\bf p},r{\bf p}'}$ which using a rough model based on the replacement of $\widetilde W_{\bf r}$ on the rectangular barrier   of the thickness $Nl_0$ and of the height $W_0$ which is around the half of ATD's gap.  This barrier couples a narrow and wide quantum wells (QWs) of widths $d_{QW}$ and $d_c$ respectively and $T_{0{\bf p},r{\bf p}'}=\delta_{{\bf p}{\bf p'}}T_{0,r}$ due to the in-plane homogeneity of the model when $\Gamma_{\bf p}$ is not depend on $\bf p$. The ground state energy of the narrow QW is ${\cal E}_0$ and the wide QW has quasi-discrete states with energies of $r$-th level $\varepsilon_r$,  moreover  ${\cal E}_0 ,\varepsilon_r\ll W_0$. The tunneling matrix element is determined by the $z$-dependent tails of wave functions, $\psi_0^>\exp (\vartheta z)$, connected to the localized state, and $\psi_{r}^<\exp [-\vartheta (z+Nl_0 )]$, connected to the $r$-th state of wide QW. Here $\vartheta \sim\sqrt{2mW_0}/\hbar$ is written through the mass of free electron, $m$, neglecting its changing in ATD. One obtains $T_{0,r}^2=W_0^2 |\psi_0^> |^2|\psi_r^< |^2\exp (-2N\vartheta l_0 )$ where $\psi_0^>$ (or $\psi_0^<$) is determined from the continuity conditions for $\psi_z$ and $d\psi_z /dz$ taken around of $z=0$ (or around of $z=-Nl_0$) for the narrow  (or wide) QW. Outside of ATD, we use the wave functions of narrow and wide QWs with the zero boundary conditions at $z=d_{QW}$ and $z=-d_c$. Within the approximation of a weak underbarrier penetration, the pre-exponential factors are written through   
\begin{equation}
\left|{\psi_0^>}\right|^2 =\frac{2{\cal E}_0}{d_{QW}W_0}, ~~~
\left|{\psi_r^<}\right|^2 =\frac{2\varepsilon_r}{d_c W_0},
\end{equation}
where $\varepsilon_r =(r\pi\hbar /d_c )^2 /2m$. Substituting Eq. (3) and $T_{0,r}^2$ into Eq. (2) and replacing $\sum_r\ldots$ by integration over energy (at $d_c\gg d_{QW}$) one obtains the leakage rate
\begin{equation}
\Gamma =\frac{4\pi{\cal E}_0}{\hbar}\left(\frac{Nl_0 }{d_{QW}} \right)^2 \exp (-2N\vartheta l_0 ) ,
\end{equation}
so that $\Gamma\propto (Nl_0{\cal E}_0 )^2\exp (-2N\vartheta l_0 )$.
\begin{figure}[t]
\vspace{0.25 cm}
\begin{center}
\includegraphics[scale=0.19]{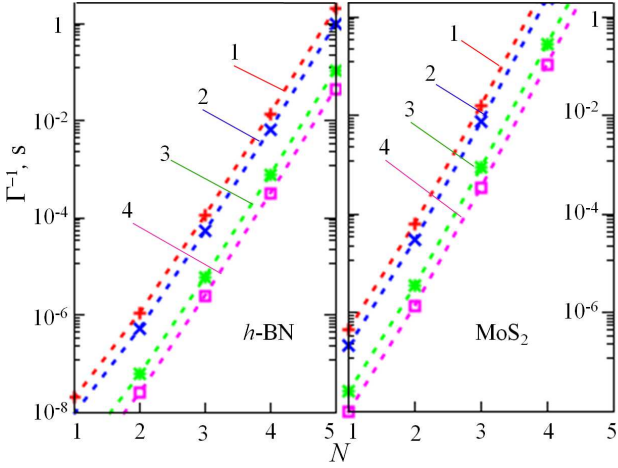}
\end{center}
\addvspace{-0.5 cm}
\caption{Leakage time, $\Gamma^{-1}$, versus number of layers, $N$, for $h$-BN (left) and MoS$_2$ (right) at ${\cal E}_0 \simeq 1.5$ (1), 2.2 (2), 6.4 (3), and 10.1 meV (4) which correspond $n_{2D}\simeq 5\times 10^8 ,~10^9 ,~5\times 10^9$, and $10^{10}$ cm$^{-2}$, respectively (see Sec. IVA). }
\end{figure}

Numerical estimates of the leakage time, $\Gamma^{-1}$, are performed with the use of a typical parameters $l_0\sim$3.2 A or $\sim$6.1 A and $W_0\sim$3 eV or 1 eV \cite{10,15,22}, when the exponential factor is determined through $2\vartheta l_0\sim$5.4 or $\sim$6.3, for $h$-BN or MoS$_2$ respectively. Calculating the pre-exponential factor with $d_{QW}$ corresponding the ground state energy ${\cal E}_0$, one obtains the dependencies of leakage time, $\Gamma^{-1}$, on $N$ shown in Fig. 2. Notice, that here ${\cal E}_0$ depends on the gate voltage or $n_{2D}$, see Table I below. According to these estimates at $n_{2D}=\simeq 5\times 10^8$ cm$^{-2}$, this time increases with $N$ from $\sim$20 ns ($N$=1) up to seconds ($N$=5) for $h$-BN or from $\sim$0.2 $\mu$s ($N$=1) up to  hour ($N$=5) for MoS$_2$. The leakage times decrease with concentration, up to two orders if $n_{2D}=10^{10}$ cm$^{-2}$. Stress one more time that a model of ATD used is oversimplified and a direct measurement of a leakage current is necessary. Up to now, there are measurements that a reflectivity of slow electrons approaches to unit (and transmissivity goes zero) \cite{23} but an accuracy of these data does not cover the $\sim$meV scale of energies considered here. As well, the microscopic calculations are not precise for this energy interval so that a further study is necessary for a quantitative description of the decay process. But all the leakage times obtained are in orders greater that the time scales determining the physics discussed below and there are no restrictions for an experimental verification of these results. Note that even the time scales below $\mu$s are interesting for some applications, e.g. for the quantum information processing.

\subsection{Image force }
Next we consider the image force induced in $N$-layer ATD placed at $0>z>-Nl_0$ and described by the longitudinal and transverse dielectric permittivities $\epsilon_{\| ,\bot}$. Since $1/\sqrt{n_{2D}}\gg 3\ell_m$, we deal with a single electron placed at $({\bf x}=0,z_0 )$. The 2D Fourier transform of the potential energy $W_{{\bf q}z}$ is governed by the Poisson equation (5a,b) with the continuity conditions for potential and its derivative (5c), taken at the ATD boundaries $z=0$ and $z=-Nl_0$:
\begin{subequations}
\begin{eqnarray}
\left(\!\frac{d^2}{dz^2} -q^2\!\right)\! W_{{\bf q}z}\! =\!
\left\{\!\! {\begin{array}{*{20}c}{-4\pi e^2\delta\left( z-z_0\right) ,} & {z\! > 0}  \\   0, & {\! - Nl_0 >z ~,}  \\
\end{array}} \right.    \\
\left[ d^2 /dz^2 -(\epsilon_\| /\epsilon_\bot )q^2\right] W_{{\bf q}z}\! =0 ~,  ~~ 0>z>-Nl_0 ~, ~~ \\
\begin{array}{*{20}c}
\left. W_{{\bf q}z} \right|_{-0}^{0}\! =\! 0,~ & \left. \frac{dW_{{\bf q}z}}{dz} \right|_{0}\!\! =\! \epsilon_\bot\!\! \left. \frac{dW_{{\bf q}z}}{dz} \right|_{-0}  ,  \\
\left. W_{{\bf q}z}\right|_{-Nl_0 -0}^{-Nl_0 +0}\!\! =\! 0 ,~ & \epsilon_\bot\!\! \left. \frac{dW_{{\bf q}z}}{dz}\right|_{-Nl_0 +0}\!\!\!\! =\!\!\left.\frac{dW_{{\bf q}z}}{dz}\! \right|_{-Nl_0 -0}  
\end{array} ~~~ 
\end{eqnarray}
\end{subequations}
and the requirements $W_{{\bf q}z\to\pm\infty}=0$. For $z>0$, the solution of this problem takes form
\begin{eqnarray}
W_{{\bf q}z}=-2\pi e^2 e^{-q|z_- |}/q +\Delta W_{qz}~,  ~~~~ \\
\Delta W_{qz}  = 2\pi e^2 \frac{e^{-qz_+}}{q}\frac{\zeta_0 \left[ 1-\exp (-q2\zeta_1 Nl_0 )\right]}{1-\zeta_0^2\exp (-q2\zeta_1 Nl_0 )}  ~, \nonumber
\end{eqnarray}
where $z_\pm =z\pm z_0$ and $\Delta W_{qz}$ is determined by the thickness of ATD, $Nl_0$, as well as the parameters $\zeta_0 =(\sqrt{\epsilon_\|\epsilon_\bot}-1)/(\sqrt{\epsilon_\|\epsilon_\bot} +1)<1$ and $\zeta_1 =\sqrt{\epsilon_\| /\epsilon_\bot}$.

In the $({\bf x},z)$-domain, the polarization-induced contribution, $\Delta W_{xz}$, is transformed into the series \cite{24}
\begin{equation}
\Delta W_{|{\bf x}|z}\!\! =\! \frac{e^2 \zeta_0}{\sqrt{{\bf x}^2\! +\! z_+^2}}\! - e^2\! \sum\limits_{n = 1}^\infty\!\!\frac{(1-\zeta_0^2 )\zeta_0^{2n-1}}{\sqrt {{\bf x}^2\! +\! (z_+\! +\! 2\zeta_1 Nl_0 n)^2}} ~.
\end{equation}
The image potential is suppressed with increasing of ${|\bf x}|$ and its maximum value at ${\bf x}=0$ is given by
\begin{equation}
\Delta W_{x=0z}\!\! =\!\frac{e^2\zeta_0}{z_+}F_{\zeta_1 Nl_0 /z_+} , ~ F_a\! =\! 1\! -\!\!\sum\limits_{n=0}^\infty \frac{(1\! -\!\zeta_0^2 )\zeta_0^{2n}}{1\! +\! 2a(n+1)} ~.
\end{equation}
For a thick dielectric, $\zeta_1 Nl_0 /z_+\gg 1$, one obtains the standard image potential $e^2\zeta_0 /z_+$ \cite{7} while $F_{a\to 0}=0$ and the image effect is negligible if $\zeta_1 Nl_0 /z_+\ll 1$. The function $F_{\zeta_1 Nl_0 /z_+}$ describing suppression of the polarization contribution in $N$-layer ATD is shown in Fig. 3. The factor $\zeta_0\sim$0.7 or $\sim$0.8 for $h$-BN or MoS$_2$ while $\zeta_1\sim$1.45 for both materials and these data are weakly dependent on $N$. \cite{15,25} The contribution of $\Delta W$ into Eq. (8) is negligible under the condition $F_{\zeta_1 Nl_0/z_+}\ll z_+ /z_-$. Taking $z_{\pm}\sim (3\ell_m\pm z_0 )$, where the density-dependent thickness of 2D-layer $3\ell_m$ is given in Table I below, one obtains that a contribution of $\Delta W_{xz}$ does not exceed 15\% or 20\% at $N\leq 5$ for $h$-BN or MoS$_2$, respectively. Thus, effect of a few-layer ($ N\leq 5$) ATD on 2D electrons can be modeled as a thin nontransparent barrier with a negligible polarization-induced potential. 
\begin{figure}[t]
\vspace{0.25 cm}
\begin{center}
\includegraphics[scale=0.185]{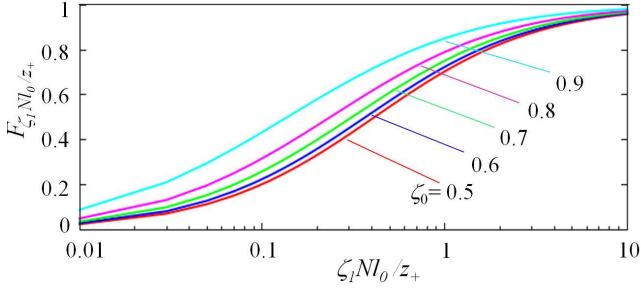}
\end{center}
\addvspace{-0.5 cm}
\caption{Factor $F_{\zeta_1 Nl_0 /z_+}$ in Eq. (8) which describes the polarization-induced contribution for different $\zeta_0$ (marked). }
\end{figure}

\section{Bending and vibrations of ATD }
Bending of the ATD, which is suspended over the long trench of width $D$ and depth $d_c$, is due to the Coulomb attraction between the 2D electrons above ATD and the back gate. This bending and vibrations of ATD are described by the in-plane and out-of-plane displacements, ${\bf u}_{{\bf x}t}$ and $z_{{\bf x}t}$, which determine the total energy \cite{18,19}
\begin{eqnarray}
E_t\!\! =\!\frac{\sigma}{2}\!\!\int\!\!{d{\bf x}}\! \left[\! {\left(\frac{\partial z_{{\bf x}t}}{\partial t}\right)^2\!\!\! +\!\!\left( {\frac{{\partial {\bf u}_{{\bf x}t} }}{{\partial t}}} \right)^2 } \right]\! +\!\!\frac{1}{2}\!\!\int\!\!{d{\bf x}}\!\left[\kappa\!\left(\Delta_2 z_{{\bf x}t}\right)^2  \right. \nonumber  \\
+ \left. 2\mu \sum\limits_{\alpha \beta } {\left( {{\widetilde u}_{\alpha \beta } } \right)^2 }  + \lambda \sum\limits_\alpha  {\left( {{\widetilde u}_{\alpha \alpha } } \right)^2}\right] -\!\int {d{\bf x}} p_{{\bf x}t} z_{{\bf x}t} ~, ~~~~~ \\
{\widetilde u}_{\alpha\beta}  = \frac{1}{2}\left( {\frac{{\partial u^{(\alpha )}_{{\bf x}t} }}{\partial x_\beta} + \frac{{\partial u^{(\beta )}_{{\bf x}t} }}{{\partial x_\alpha  }} + \frac{{\partial z_{{\bf x}t} }}{{\partial x_\alpha  }}\frac{{\partial z_{{\bf x}t} }}{{\partial x_\beta  }}} \right) ~ . ~~~~~ \nonumber
\end{eqnarray}
The elastic properties of ATD are characterized by the Lame parameters, $\mu$ and $\lambda$, the bending stiffness, $\kappa$, and the 2D density of mass, $\sigma$. Also $p_{{\bf x}t}$ describes pressure of 2D electrons on ATD and ${\widetilde u}_{\alpha\beta}$ is the strain tensor. Further, by varying $E_t$, we obtain the equations of motion for these displacements from which we determine the steady-state profile of the suspended ATD and the dispersion law for flexural vibrations. 

Considering the case of ATD with the  edges, clamped at $x=\pm D/2$,  for the steady-state regime one deals with the $x$-dependent displacements $(u^{(x)}_x,0,z_x)$ which are governed by the time-independent system of equations \cite{19,26}
\begin{subequations}
\begin{eqnarray}
\kappa\frac{d^4 z_{x}}{dx^4}-U(2\mu +\lambda )\frac{d^2 z_{x}}{dx^2}=p_\bot , ~~~~~~~~~  \\   
\frac{du^{(x)}_{x}}{dx}+\!\frac{1}{2}\!\left(\frac{dz_{x}}{dx}\right)^2\!\! = \! U\! =\!\int_{-D/2}^{D/2}\frac{dx}{2D}\!\left(\frac{dz_x}{dx}\right)^2 . 
\end{eqnarray}
\end{subequations}
Here $p_\bot\approx -2\pi (en_{2D})^2$ is the steady-state part of the transverse pressure, \cite{9} $U$ is the $xx$-component of strain, which is written through $dz_x /dx$ [the right-hand part of Eq. (10b)] after the integration across the trench. The out-of-plane displacement $z_x$ is obtained from Eq. (10a) through $U$ which is determined from the self-consistency condition given by Eq. (10b). These results take form:
\begin{equation}
z_x\! =\!\frac{p_\bot (D^2 /4-x^2 )}{2U(2\mu +\lambda )}\! +\delta z_x , ~~  U\!\approx\!\frac{1}{24}\!\left[\frac{p_\bot  D}{(2\mu +\lambda )U} \right]^{2} ,
\end{equation}
where $\delta z_x$ stands for the correction at edges to $z_x$ localized at $|x \pm D/2| \le\sqrt{\kappa /U(2\mu +\lambda)}$ and the strain is $U\!\approx\! [|p_\bot |D/\sqrt{24}(2\mu +\lambda )]^{2/3}$. This correction gives  negligible contributions to the integral in (10b) and to the coefficients (14) below. At the width $D=$10 $\mu$m and the pressure $p_\bot$ corresponding to $n_{2D}=10^{10}$ cm$^{-2}$, one obtains $U\sim 1.9\times 10^{-5}$ or $3.3\times 10^{-5}$ and the maximal bending of ATD, $z_{x=0}\propto (n_{2D}D^2 )^{2/3}$ is estimated as $\sim 2.7$ A or $\sim 3.5$ A for the parameters of $h$-BN or MoS$_2$, respectively. The  bending $z_{x=0}$ do not depend on $\kappa$.  Here and below we use the typical Lame parameters of $h$-BN (or MoS$_2$): $\mu\sim 1.2\times 10^5$ dyn/cm (or $\sim 5\times 10^4$ dyn/cm) and $\lambda\sim 10^5$ dyn/cm (or $\sim 4.3\times 10^4$ dyn/cm) which are weakly dependent on $N$. \cite{14, 27, 28} Because of $z_{x=0}$ is negligible in comparison to all the sizes under consideration (3$\ell_m$, $d_c$, and $D$), the curvature of ATD is only taken into account under the description of the vibrations in this section, while further (Sec. IV and V) we use the flat capacitor approximation.

Vibrations of the suspended ATD are described by the weak contributions to the in-plane and out-of-plane displacements, $\delta{\bf u}_{{\bf x}t}$ and $\delta z_{{\bf x}t}$, which are governed by the linearized system of equations:
\begin{subequations}
\begin{eqnarray}
\begin{array}{*{20}c}
\left(\!\sigma\!\frac{\partial^2}{\partial t^2}\! +\!\kappa \Delta_2^2\right)\!\delta z_{{\bf x}t}\! -\!\!\left({2\mu\! +\!\lambda}\right)\!\!\left\{\! \frac{\partial }{\partial x}\!\left[\!\left(\! {\frac{dz_x }{dx}}\! \right)^2\!\!\!\frac{\partial }{\partial x}\!\right]\!\! +\! U\!\frac{\partial ^2}{\partial x^2}\! \right\}\!\delta z_{{\bf x}t}   \\ 
\! =\! (2\mu\! +\!\lambda )\!\frac{\partial }{\partial x}\!\left(\! \frac{dz_x}{dx}\!\frac{\partial \delta u_{{\bf x}t}^{(x)} }{\partial x}\! \right)\! +\!\mu\frac{dz_x}{dx}\!\left(\!\frac{\partial^2\!\delta u_{{\bf x}t}^{(x)}}{\partial x^2}\! +\!\frac{\partial^2 \delta u_{{\bf x}t}^{(y)}}{\partial x\partial y}\! \right) , \end{array}  ~~~ \\ 
\begin{array}{*{20}c}\left(\sigma\frac{\partial^2}{\partial t^2}-\widehat{\cal M}\right)\left| \begin{array}{*{20}c} {\delta u_{{\bf x}t}^{(x)} }  \\
{\delta u_{{\bf x}t}^{(y)}}\end{array}\right| ~~~~~~~~ \\
=\!\!\left\{\!\! {\begin{array}{*{20}c}\!
{\mu\!\left[{2\frac{\partial }{{\partial x}}\!\!\left( {\frac{{dz_x }}{{dx}}\frac{{\partial \delta z_{{\bf x}t} }}{{\partial x}}}\!\right)\! +\! \frac{{dz_x }}{{dx}}\frac{{\partial ^2 \delta z_{{\bf x}t} }}{{\partial y^2 }}}\right]\!\! +\!\lambda\frac{\partial}{\partial x}\!\!\left( {\frac{{dz_x }}{dx}\frac{{\partial \delta z_{{\bf x}t} }}{{\partial x}}} \right)}   \\
 \!\!\!\!\!\!\!\!{\mu \frac{\partial }{{\partial x}}\left( {\frac{{dz_x }}{{dx}}\frac{{\partial \delta z_{{\bf x}t}}}{\partial y}}\right)} ~~~~~~~    \\
\end{array}} \right. , ~~  
\end{array}  
\end{eqnarray}
\end{subequations}
where the lateral vibrations are described by the operator
\begin{equation}
\widehat{\cal M}\!\equiv\!\left|\! \begin{array}{*{20}c}
\mu\! \left(\! 2\frac{\partial^2}{\partial x^2}\! +\!\frac{\partial^2}{\partial y^2}\!\! \right)\!\!-\!\!\lambda\frac{\partial ^2}{\partial x^2}\!\!\!\! & \mu\frac{\partial^2}{\partial x\partial y}  \\
   \mu \frac{\partial ^2}{\partial x\partial y} & \!\!\!\! \mu\!\left(\!2\frac{\partial^2}{\partial y^2}\!\! +\! \frac{\partial ^2}{\partial x^2 }\!\! \right)\!\! -\!\!\lambda\frac{\partial ^2}{\partial y^2}  \end{array}\!\right| . 
\end{equation}
Further, we perform the Fourier transforms of the displacements so that $(\delta{\bf u}_{{\bf x}t},\delta z_{{\bf x}t})=\exp (iq_y y-i\omega t)\sum_k e^{-iq_k x}(\delta{\bf u}_{kq_y\omega},\delta z_{kq_y\omega})$, where $q_k =2\pi k/D$ and $q_y$ are the components of the in-plane wave vector. For the short wavelength region, the out-of-plane vibrations are determined by the $\Delta_2^2\delta z_{{\bf x}t}$ contribution into Eq. (12a) and there is the quadratic dispersion law for the flexural vibrations, $\omega_{q_k ,q_y}\simeq\sqrt{\kappa /\sigma}(q_k^2 +q_y^2)$. \cite{17,18,19} For the long wavelengths, we consider the system (12) for $(\delta{\bf u}_{kq_y\omega},\delta z_{kq_y\omega})$ in the region $\kappa (q_k^4 ,q_y^4 )/\sigma\ll\omega^2\ll (2\mu +\lambda )(q_k^2 ,q_y^2 )/\sigma$, when the dispersion law becomes linear one. If $q_k D,q_y D\gg 1$, this system takes form 
\begin{subequations}
\begin{eqnarray}
\begin{array}{*{20}c}
\left( {\frac{\omega }{{\omega _f }}} \right)^2 \delta z_{kq_y\omega}- \sum\limits_{k'=-\infty }^\infty  {A_{kk'} } \delta z_{k'q_y\omega}  ~~~  \\
 = \frac{{L_f }}{D}\sum\limits_{k' =  - \infty }^\infty  {\left( {a_{kk'} \delta u_{k'q_y \omega }^{(x)}  - i\chi q_y Db_{kk'} \delta u_{k'q_y \omega }^{(y)} } \right)} , \end{array}  \\
\begin{array}{*{20}c}\left| {\begin{array}{*{20}c}
   {q_k^2  + \chi q_y^2 } & {\chi q_k q_y }  \\
   {\chi q_k q_y } & {q_y^2  + \chi q_k^2 }  \\
\end{array}} \right|\left| {\begin{array}{*{20}c}
   {\delta u_{kq_y \omega }^{(x)} }  \\
   {\delta u_{kq_y \omega }^{(y)} }  \end{array}} \right|  ~~~~~  \\
  = \frac{1}{{L_f D}}\sum\limits_{k' =  - \infty }^\infty  {\left| {\begin{array}{*{20}c} {-a_{kk'} \delta z_{k'q_y \omega } }  \\
   {i\chi q_y Db_{kk'} \delta z_{k'q_y \omega } }  \\
\end{array}} \right|} . \end{array}
\end{eqnarray}
\end{subequations}
Here $\chi\equiv\mu /(2\mu +\lambda )\sim 0.36$ for the both ATD under consideration and we have introduced the characteristic length and frequency,  $L_f =U(2\mu +\lambda )/|p_\bot |$ and $\omega_f =L_f^{-1}\sqrt{(2\mu +\lambda )/\sigma}$. There is no time delay between $\delta{\bf u}$ and $\delta z$ in Eq. (14b) and the coefficients in Eqs. (14) are written through $\Delta k=k-k'$ as follows   
\begin{equation}
\begin{array}{*{20}c}
~~~ A_{kk'}\!\approx\!\left\{\begin{array}{*{20}c}{(q_k D)^2 ,} & {k=k'}  \\
 {\frac{{(q_k D)^2 +\chi (q_y D)^2}}{2(\pi \Delta k)^2},} & {\Delta k\ne 0} ~~~ \end{array} , \right.  \\
a_{kk'}\!\approx\! i\frac{{(q_{k'}D)^2 -\chi (q_y D)^2}}{(2\pi\Delta k)^2}, ~ b_{kk'}\!\approx\! -\frac{{q_{k'}D}}{2\pi\Delta k}, ~\Delta k\ne 0  \\
\end{array} 
\end{equation}
and $a_{kk}=b_{kk}=0$. We dropped out the factors $(-1)^{\Delta k}$ from these coefficients because the system (14) does not changed after the simultaneous replacing the displacements $(\delta{\bf u}_{kq_y\omega},\delta z_{kq_y\omega})$ by $(-1)^k (\delta{\bf u}_{kq_y\omega},\delta z_{kq_y\omega})$.

Eliminating $\delta{\bf u}$ from Eqs. (14) one obtains the closed equation for the out-of-plane displacement:  
\begin{equation}
\begin{array}{*{20}c}
 [ (\omega /\omega_f\! )^2\!\! -\!\!{\cal A}_{\bf q}]\delta z_{kq_y\omega}\!\! - \!\!\!\!\sum\limits_{\Delta k\ne 0}\!\!\!\!{\cal K}_{k\Delta kq_y}\!\delta z_{k+\Delta kq_y\omega}\! =\! 0 ,  \\
  {\cal A}_{\bf q}=[13(q_k D)^2 -\chi (q_y D)^2 ]/12 , \\
\end{array} 
\end{equation}
where ${\cal K}_{k\Delta kq_y}\propto (q_{k,y}D)^2$ appears due to the contribution of the in-plane vibrations and this kernel rapidly decreases with growth of $\Delta k$. The equation (16) should be solved with boundary conditions at $x=\pm D/2$ that provide thermalization of the suspended ATD. For slowly varying displacements, when $k\gg\Delta k$ and $\delta z_{k+\Delta kq_y\omega}\approx\delta z_{kq_y\omega}$ one obtains the dispersion relation in the form: $(\omega /\omega_f )^2\approx{\cal A}_{\bf q}+\sum\nolimits_{\Delta k}{\cal K}_{k\Delta kq_y}$. Using ${\cal K}_{k\Delta kq_y}$ determined from the Eqs. (14), (15) and performing straightforward summations over $\Delta k$, one transforms this relation into $(\omega /\omega_f )^2\approx (qD)^2\Psi_\phi$ where factor $\Psi_\phi$ depends on the polar angle of $\bf q$. Within an accuracy $\sim$5\% this angle dependency can be approximated as $\Psi_\phi\approx (7/6)\cos^2\phi +0.075\sin^2\phi$, so that the dispersion law is anisotropic.

Finally, we connect the long and short wavelength regions and use below the phonon dispersion law  
\begin{equation}
\omega_{\bf q}\approx\sqrt{\frac{\kappa}{\sigma}q^4 +(s_x q_x )^2 +(s_y q_y )^2} \equiv q\sqrt{\frac{\kappa}{\sigma}q^2 +s_\phi^2} ~, 
\end{equation}
written through the anisotropic sound velocity, $s_\phi =\omega_f D\sqrt{\Psi_\phi}$ or through $s_{x,y}=s_{\phi =0,\pi /2}$; see similar result for graphene in Refs. 29. A conversion from the linear to quadratic dispersion law takes place at the wave vector $q_\phi\sim s_\phi\sqrt{\sigma /\kappa}$. The dispersion relation $\omega_k$ is determined by the ratio $\kappa /\sigma$, which is $\sim 1.9\times 10^7$ eV$\times$cm$^2$/g or $\sim 2.9\times 10^7$ eV$\times$cm$^2$/g for $h$-BN or MoS$_2$ respectively, and by the velocities $s_{x,y}=\omega_f D\sqrt{\Psi_{\phi =0,\pi /2}}$. Here the characteristic velocity $\omega_f D\propto\sqrt[3]{n_{2D}^2 D}$ is $\sim 5\times 10^3$ cm/s or $\sim 2\times 10^3$ cm/s and the characteristic wave vector $q_{\phi}/\Psi_\phi$ is $\sim 1.3\times 10^6$ cm$^{-1}$ or $\sim 0.4\times 10^6$ cm$^{-1}$ for the $h$-BN or MoS$_2$ at $n_{2D}=10^{10}$ cm$^{-2}$ and $D=10~\mu$m. 

The above estimates are performed  for a single-layer ATD with the use of typical bending stiffness for $h$-BN or MoS$_2$, $\kappa\sim$1.3 eV or $\kappa\sim$9 eV which are closely to the data from \cite{14, 30}, see also the references therein. The 2D density of mass, $\sigma\sim 6.7\times 10^{-8}$ g/cm$^2$ or $\sim 3.1\times 10^{-7}$ g/cm$^2$, is estimated from the bulk densities and the lattice constants, see similar calculations in  \cite{31}. In the case of $N$-layer ATD, $\sqrt{\kappa /\sigma}$ increases slowly, $\propto N^{(\theta -1)/2}$ with $\theta\sim 2.3$, because of $\sigma\propto N$ and of the relation $\kappa\propto N^{\theta}$ \cite{32,33} (at $N\gg 1$ the stronger dependency $\kappa\sim N^3$ takes place). As a result, $s_\phi \propto 1/\sqrt{N}$ and $q_{\phi}$ decreases as $\propto 1/N^{\theta /2}$. Since the typical wave vector of 2D-electrons, $q_T =\sqrt{2mT}/\hbar$, is $\sim 1.6\times 10^6$ cm$^{-1}$ at $T=$1 K, the transition between $\omega_{\bf q}\propto q$ and $\omega_{\bf q}\propto q^2$ appears at the temperature range under consideration, depending on the parameters of device ($n_{2D}$, $D$, and $N$). 

\section{2D spectrum and scattering of electrons}
Now we consider the confined electronic states and  describe the energy diagram (the ground level and ionization energies) within the self-consistent approach. The scattering processes of 2D electrons are analyzed for the cases of interaction  with the  flexural phonons (Sec. III) or with the roughness of ATD caused by a small-size monolayer islands. 

\subsection{Self-consistent energy spectrum } 
Neglecting the polarizability of the ATD placed at $z=0$ we use the boundary condition $\psi_{z\to 0}=0$ and the system of the $z$-dependent Schrodinger and Poisson equations takes form:
\begin{subequations}
\begin{eqnarray}
\left(\widehat{p}_z^2 /2m+W_{z}-{\cal E}\right)\psi_{z}=0, ~~ z>0 ,  \\
d^2 W_{z}/dz^2 =-4\pi e^2 n_{2D}\psi_{z}^2 , ~~ z>-d_c .  
\end{eqnarray}
\end{subequations}
The wave function $\psi_{z}$ is normalized by the condition $\int_0^{\infty}{dz}\psi_{z}^2 =1$ and the potential energy $W_z$ is satisfied by the boundary condition at gate $W_{z=-d_c}=eV_c$ written through the bias voltage, $V_c$, and the charge of electron, $e$. Using the continuity conditions for $W_z$ and $dW_z /dz$ at $z=0$, one obtains the solution of (18b) in the form:
\begin{equation}
\frac{W_z -eV_c}{4\pi e^2 n_{2D}}\! =\!\left\{\!\!\!\!\!\begin{array}{*{20}c} \! d_c +z-\!\int_{0}^z{dz'}(z\! -\! z')\psi _{z'}^2  , ~ {z>0}  \\
 ~~~~~~~~ (z + d_c )~, ~~~~~~~~ {-d_c <z<0} \\   \end{array} \right. .
\end{equation}
Below we choose the zero-point energy at the ATD position, $W_{z=0}=0$, so that $n_{2D}$ and $V_c$ are connected as follows: $4\pi e^2 n_{2D}=|e|V_c /d_c$.   

We search the variational solution of Eq. (18a) with the trial wave function $\psi_{z}=z\exp [-z/(2\ell )]/\sqrt{2\ell^3}$ dependent on the characteristic length $\ell$. The energy functional takes form ${\cal E}_{\ell}=(\hbar /2\ell )^2/2m+(1/2)\int_0^{\infty}dz W_{z}\psi_{z}^2$ and after the straightforward integrations one obtains \cite{1,34}
\begin{equation}
{\cal E}_{\ell}=\frac{(\hbar /\ell )^2}{8m}+\frac{33|e|V_c\ell}{32(d_c +3\ell )}
\end{equation}
with the minimum at $\ell =\ell_m$. We restrict ourselves by the case of the wide plane capacitor, $d_c\gg 3\ell_m$, when $n_{2D}\simeq V_c /4\pi |e|d_c$ (i.e. $n_{2D}$ is determined by the electric field applied to the device, $E_c\equiv V_c /d_c$) and the explicit expressions for the thickness of the electron layer, $3\ell_m$, and the ground state energy, ${\cal E}_0$, are: 
\begin{equation}
3\ell_m \simeq \sqrt[3]{\frac{54a_B}{33\pi n_{2D}}}, ~~~ {\cal E}_0 \simeq \frac{5}{4}{\cal E}_R \left( {\frac{33\pi}{2}n_{2D} a_B^2 }\right)^{2/3} 
\end{equation} 
where ${\cal E}_R =e^2 /2a_B$ is the Rydberg energy. In addition, using $W_{z\to\infty}=12\pi e^2 n_{2D}\ell_m$ we estimate the energy of ionization, ${\cal E}_i =W_{z\to\infty}-{\cal E}_0$, as ${\cal E}_i\approx 0.17{\cal E}_0$.
\begin{table}[t]
\vspace{0.3 cm}
	\centering
		\begin{tabular}{|c|c|c|c|c|}  \hline
		$n_{2D}$, cm$^{-2}$ & $E_c$, kV/cm  & $3\ell_m$, A	& ${\cal E}_0$, meV  & ${\cal E}_i$, meV  \\  \hline
		$5\times 10^8$  &  0.9  &  175 & 1.5 &  0.24  \\  
		$10^9$  & 1.8  & 140  &  2.2  &  0.36   \\ 
		$5\times 10^9$  & 9  & 82  &  6.4  &  1   \\ 
		$10^{10}$  & 18  &  65  &  10.1  &  1.65  \\  	\hline	
	\end{tabular}
	\caption{Energies ${\cal E}_0$ and ${\cal E}_i$ and thickness of 2D layer $3\ell_m$ shown in Fig. 1 versus the applied fields, $E_c =V_c /d_c$, or 2D concentrations, $n_{2D}$.}
\end{table}

Thus, the parameters of electronic state are determined by $n_{2D}$ or 
$E_c$: if $E_c$ varies from $\sim$1 kV/cm to $\sim$18 kV/cm, the thickness of layer, $3\ell_m\propto E_c^{-1/3}$, decreases in $\sim$2.5 times and the ground state energy, ${\cal E}_0\propto E_c^{2/3}$, increases in $\sim$7 times, see Table 1. \cite{35} The correspondent energies of ionization are between 0.25 meV and 1.65 meV, so that the regime of transverse localization takes place in the low temperature region, which are below $\sim$2 K or $\sim$15 K, for low or high concentration. For $E_c\sim 1\div 18$ kV/cm, the Fermi energies, $n_{2D}/\rho_{2D}$, are between  1.2$\div$24 $\mu$eV ($\rho_{2D}=m/\pi\hbar^2$) and electrons of any concentration are nondegenerate if $T$ exceeds $\sim$0.1 K. The typical interaction energy between electrons, $\varepsilon_C =e^2\sqrt{\pi n_{2D}} \propto\sqrt{E_c}$ varies on the interval 6$\div$25 meV and exceeds the kinetic energy in tens time but it is far from the Wigner crystallization condition, $\varepsilon_C /{T}\geq 140$. \cite{7} Further, we restrict our consideration by the case of the nondegenerate nonideal plasma (the Boltzmann liquid regime).

\subsection{Scattering via  flexural phonons }
The effective energy of the flexural vibrations, which correspond to the approach given by Eqs. (16) and (17), is written as
\begin{eqnarray}
\delta E_t  =\frac{1}{2}\int_{(L^2 )} {d{\bf x}} \left[\frac{\delta p_{{\bf x}t}^2}{\sigma}+ \kappa\left(\Delta_2\delta z_{{\bf x}t}\right)^2\right. \\
\left. +\left( s_x\frac{\partial}{\partial x}\right)^2\delta z_{{\bf x}t}+\left( s_y\frac{\partial}{\partial y}\right)^2\delta z_{{\bf x}t}\right] ,
\nonumber
\end{eqnarray}
where $\delta p_{{\bf x}t}=\sigma\partial\delta z_{{\bf x}t}/\partial t$ is the density of momentum and $L^2$ is a normalization area. Under the standard procedure of canonical quantization $\delta z_{{\bf x}t}$ is replaced by  the transverse displacement operator $\widehat{\delta z}_{\bf x}$ given by
\begin{equation}
\widehat{\delta z}_{\bf x} = L^{-1}\sum\nolimits_{\bf q} \sqrt {\frac{\hbar}{2\sigma\omega _{\bf q}} } e^{i{\bf q}\cdot{\bf x}}\widehat{b}_{\bf q} + H.c. 
\end{equation}
and Eq. (22) is transformed into the Hamiltonian $\widehat H_{ph} =\hbar \sum\nolimits_{\bf q} {\omega_{\bf q}}(\widehat{b}_{\bf q}^+\widehat{b}_{\bf q}+1/2)$. Here $\widehat{b}_{\bf q}^+$ and $\widehat{b}_{\bf q}$ are the creation and annihilation operators for the flexural phonon with the wave vector $\bf q$ and frequency $\omega_{\bf q}$ given by Eq. (17). 

Further, we derive the transition probabilities between 2D states with momenta $\bf p$ and ${\bf p}'$ caused by the interaction of 2D electrons with the flexural phonons. The effect of the vibration-induced curvature of ATD is taken into account by the use of the zero boundary condition for $\psi_z$ at the surface $z=\delta z_{\bf x}$. We perform the unitary transformation $\exp (-i\delta z_{\bf x}\widehat{p}_z /\hbar )$ of Eq. (18a) written for the region $z\geq\delta z_{\bf x}$, which shifts the electron coordinate so that it is counted off from the flat surface $z=0$. \cite{36} Using the operator (23) and remaining $\propto\widehat\delta z$ contributions one obtains the operator of the electron-phonon coupling:
\begin{equation}
\widehat{\delta H}_{e,ph}\! =\!\widehat{\delta z}_{\bf x} \frac{dW_z}{dz}\! +\!\left(\! \bnabla_{\bf x}\widehat{\delta z}_{\bf x}\cdot \widehat{\bf p}_{\bf x}\! -\!\frac{i\hbar}{2}\nabla_{\bf x}^2\widehat{\delta z}_{\bf x} \right)\frac{\widehat{p}_z}{m} .
\end{equation}
The first and second addenda here are due to modulation of the potential and kinetic energies, respectively. The kinetic part of the coupling energy (which is $\propto\widehat{p}_z /m$) gives zero contribution under the averaging over the ground state. For the case of the in-plane transport of 2D electrons, Eq. (24) is transformed into 
\begin{equation}
\widehat{\delta{\cal H}}_{e,ph}\! =f_\bot\widehat{\delta z}_{\bf x} , ~ 
f_\bot\!\! =\!\int_0^\infty\!\! {dz}\frac{dW_z}{dz}\psi_z^2 =2\pi e^2 n_{2D}
\end{equation}
and the coupling strength, $f_\bot\propto E_c$, was calculated here with the use the trial wave function and $dW_z /dz$ from Sect. IVA. 

The  interaction due to emission and absorption of the flexural phonons  is described by Eqs. (23) and (25) and we obtain the transition probability from the electronic state $\bf p$ into ${\bf p}'$ one as follows
\begin{eqnarray}
{\cal W}_{{\bf p},{\bf p}'}^{(ph)}  = \frac{\pi f_\bot^2}{L^2\sigma \omega_{\bf q}}\left[ (N_{\bf q}+1)\delta (\varepsilon_p -\varepsilon_{p'}  + \hbar\omega_{\bf q} ) \right.  \nonumber \\
\left. + N_{\bf q} \delta (\varepsilon_p -\varepsilon_{p'} -\hbar\omega_{\bf q}) \right]_{\hbar{\bf q}={\bf p}-{\bf p'}} .
\end{eqnarray}
Here $N_{\bf q}$ is the Planck distribution of the flexural phonons at the equilibrium temperature $T$. The transition from ${\bf p}'$ into $\bf p$ is determined through the detailed equilibrium condition ${\cal W}_{{\bf p}',{\bf p}}^{(ph)}={\cal W}_{{\bf p},{\bf p}'}^{(ph)}\exp [(\varepsilon_p -\varepsilon_{p'})/T]$. Because the momentum transfer is of the order of the equilibrium momentum, $p_T\simeq  \sqrt{2mT}$, a typical energy of emitted and absorbed phonons is $\sim \hbar\omega_{p_T /\hbar} =\sqrt {\kappa /\sigma}p_T^2 /\hbar\sim 10^{-2}T$ both for $h$-BN and for MoS$_2$. Due to the weakness of the energy transfer under the phonon-induced scattering,  the quasielastic approximation is valid and Eq. (26) is written as $\overline{\cal W}_{{\bf p},{\bf p}'}+\Delta{\cal W}_{{\bf p},{\bf p}'}$:
\begin{subequations}
\begin{eqnarray}
\overline{\cal W}_{{\bf p},{\bf p}'}\simeq \frac{2\pi f_\bot^2 T}{L^2 \sigma\hbar\omega_{({\bf p}-{\bf p}')/\hbar}^2}\delta (\varepsilon_p -\varepsilon_{p'}) ~, ~~~~~  \\
\Delta{\cal W}_{{\bf p},{\bf p}'}\simeq\frac{\pi f_\bot^2\hbar}{L^2\sigma}\left[\delta '(\varepsilon_p -\varepsilon_{p'})+T\delta ''(\varepsilon_p -\varepsilon_{p'}) \right] .  
\end{eqnarray}
\end{subequations}
The elastic probability $\overline{\cal W}_{{\bf p},{\bf p}'}$ is $\propto\omega_{\bf q}^{-2}$ while the non-elastic part of the phonon-induced scattering $\Delta{\cal W}_{{\bf p},{\bf p}'}$ does not depend on $\omega_{\bf q}$.

\subsection{Scattering by monolayer roughness of ATD } 
In addition to the scattering via phonons, relaxation can be caused by the monolayer islands which form a rough boundary of ATD described by the steady-state displacement ${\delta z}_{\bf x}$. The interaction of 2D electron with these islands is described similarly to Eq. (25): 
\begin{equation}
\widehat{\delta{\cal H}}_{e,rh} =f_\bot\sum\nolimits_{(j)}{\delta z}_{\bf x}^{(j)} , ~~ j=1,\ldots ,N_{is} ~,
\end{equation}
where $N_{is}/L^2\equiv n_{is}$ is the concentration of islands and $\delta z_{\bf x}^{(j)}$ describes the $j$-th scatterer placed at a random position ${\bf x}_j$. We consider the model of an identical islands, when the transition probability between the states $\bf p$ and ${\bf p'}$  is given by the standard expression \cite{20}
\begin{equation}
{\cal W}_{{\bf p},{\bf p}'}^{(rh)}=\frac{2\pi f_\bot^2 n_{is}}{\hbar L^2 }
|{\delta z}_{\bf k}|^2\delta (\varepsilon_p -\varepsilon_{p'})|_{\hbar {\bf k} ={\bf p}-{\bf p}'} .
\end{equation}
Here the momentum conservation law $\hbar{\bf k}={\bf p}-{\bf p}'$ is taken into account and the Fourier transform of the form-factor $\delta z_{\bf x}$ is performed. For an  island of the disk shape with radius $r_{is}$, which is placed at ${\bf x}_j =0$ so that $\delta z_{\bf x}=l_0$ if $|{\bf x}|<r_{is}$ and $\delta z_{\bf x}=0$ if $|{\bf x}|>r_{is}$, one obtains the isotropic form-factor
\begin{equation}
 {\delta z}_k =\!\int\!{d{\bf x}} e^{ - i{\bf kx}} \delta z_{\bf x}  = \frac{{l_0 r_{is} }}{k}J_1 \left( {kr_{is} } \right) ,
\end{equation}
which is written through the Bessel function of the 1st order, $J_1(\ldots )$.

For such a model, the transition probability takes form
\begin{equation}
 {\cal W}_{{\bf p},{\bf p}'}^{(rh)}\! =\!\frac{2\eta_{is}}{\hbar L^2}\! \left[\!{\frac{f_\bot  l_0}{{\Delta p/\hbar }} J_1\!\left(\! {\frac{\Delta pr_{is}}{\hbar }}\!\right)}\!\right]^2\!\!\delta \left( {\varepsilon_p\! -\!\varepsilon_{{\bf p} - \Delta {\bf p}} } \right)
\end{equation}
where $\Delta{\bf p}={\bf p}-{\bf p}'$ means the momentum transfer and $\eta_{is}=n_{is}\pi r_{is}^2\ll 1$ determines part of ATD covered by islands (taking $r_{is}\sim 50$ A and $n_{is}\sim 10^9$ cm$^{-2}$ one obtains $\eta_{is}\sim 8\times 10^{-4}$). An efficiency of scattering via roughness is determined by their characteristics, $l_0$, $r_{is}$, and $\eta_{is}$, but does not depend on any other parameters of ATD. This is an elastic process and, similarly to Eq. (27a), the probability (31) is written as the $\delta$-function multiplied by the $\Delta p$-dependent prefactor. For $ph$-scattering the prefactor is divergent as $\Delta p^{-2}$ at $\Delta p\to 0$ and decreases as $\Delta p^{-4}$ if $\Delta p\gg\hbar q_\phi$, see Eq. (27a). For $rh$-scattering the prefactor is constant at $\Delta p\to 0$ and there is a non-monotonic  decreasing due to the contribution of the Bessel function at $\Delta p\geq\hbar /r_{is}$. For the case of non-identical islands of arbitrary shape, it is necessary to carry out a more complex averaging and replace the form-factor (30). But the result is again similar to Eq. (31) and is expressed in terms of the concentration and the characteristic size of islands, which determine the magnitude of ${\cal W}_{{\bf p},{\bf p}'}^{(rh)}$ and its cutoff with increasing of $\Delta p$.

\section{In-plane electron transport }
Next, we consider the in-plane transport limited by the relaxation processes discussed in Sect. IV. Because of the strong electron-electron interaction we employ the shifted quasi-equilibrium distribution, $f_{\bf p}=\widetilde f_\varepsilon +\Delta f_{\bf p}$, characterized by the electron temperature $T_e$ and the drift velocity ${\bf v}_{dr}$. Here $\widetilde f_\varepsilon$ is the Boltzmann distribution at temperature $T_e$ and the weak anisotropic contribution takes form $\Delta f_{\bf p}=({\bf v}_{dr }\cdot{\bf p})\widetilde f_\varepsilon /T_e$. The losses of the drift velocity and energy per electron, ${\bf R}_{dr}$ and $Q$, are introduced by the relations \cite{37}
\begin{subequations}
\begin{eqnarray}
{\bf R}_{dr}\! =\!\sum\limits_{{\bf p},{\bf p}'}\! {\frac{{{\bf p} - {\bf p}'}}{mn_{2D} L^2}\left[ {\overline{\cal W}_{{\bf p},{\bf p}'}\! +\! {\cal W}_{{\bf p},{\bf p}'}^{(rh)} } \right]}\!\left( {\Delta f_{{\bf p}'}\! -\! \Delta f_{\bf p} }\right) , ~~~~  \\
Q\! =\!\sum\limits_{{\bf p},{\bf p}'}\!\frac{\varepsilon -\varepsilon '}{{n_{2D} L^2}}\Delta{\cal W}_{{\bf p},{\bf p}'}\! \left[ {\exp\left({\frac{\varepsilon '-\varepsilon}{T}}\right)\widetilde f_{\varepsilon '} - \widetilde f_\varepsilon  } \right] . ~~~~
\end{eqnarray}
\end{subequations}
In addition we restrict ourselves by the weak heating case, $T\gg |T_e -T|$, when ${\bf R}_{dr}$ and $Q$ are connected with the momentum and energy relaxation rates, $\nu_m$ and $\nu_e$, according to: ${\bf R}_{dr}=-\nu_m{\bf v}_{dr}$ and  $Q=-\nu_e (T_e -T)$.

\subsection{Momentum relaxation via phonons }
Substituting $\Delta f_{\bf p}$ into Eq. (32a) we obtain the momentum relaxation rate as follows
\begin{equation}
\nu_m  =\sum\limits_{{\bf p},{\bf p}'}\frac{[{\bf v}_{dr}\cdot ({\bf p}-{\bf p}')]^2}{n_{2D}L^2 Tmv_{dr}^2}\left[\overline{\cal W}_{{\bf p},{\bf p}'}+W_{{\bf p},{\bf p}'}^{(rh)}\right] f_\varepsilon ~.
\end{equation}
Here we calculate the rate $\nu _m^{(ph)}$ described the contribution of the elastic scattering via the phonons and leave in $\nu_m$ the contribution of Eq. (27a) only:
\begin{equation}
\nu _m^{(ph)}\!\! =\!\frac{2\pi f_\bot^2}{L^4\sigma\hbar\rho_{2D}}\!\!\sum\limits_{{\bf p},\Delta {\bf p}}\!\!\frac{e^{-\varepsilon /T}\!({\bf v}_{dr}\!\!\cdot\!\Delta {\bf p}\! )^2}{\omega _{\Delta{\bf p}/\hbar}^2 Tmv_{dr}^2}\delta\! \left(\varepsilon_p\!\! -\!\varepsilon_{{\bf p}-\Delta{\bf p}}\right) . 
\end{equation}
After the standard averaging over angles this equation is written through the relaxation rates along $x$- and $y$-directions, $\nu_m^{(x)}$ and $\nu_m^{(y)}$, as follows $\nu _m^{(ph)}=\nu_m^{(x)}\cos^2\widehat{{\bf v}_{dr},{\bf e}_x}+\nu_m^{(y)}\sin^2\widehat{{\bf v}_{dr},{\bf e}_x}$. Performing the simple integration over $\varepsilon_p$ one transforms these rates into the double integrals
\begin{eqnarray}
\nu _m^{(\alpha )}\!\! =\!\widetilde{\nu}_{ph}\!\!\!\int\limits_0^\infty\! \frac{d\Delta x}{\sqrt {\Delta x}} \Phi_{g\Delta x}^{(\alpha )} e^{-\Delta x/4}\! , ~ \widetilde{\nu}_{ph}\!\! =\!\frac{\sqrt{\pi}f_\bot^2}{16\sigma (\omega _f D)^2 \hbar} , ~~~~ \\
\|\Phi_{g\Delta x}^{(x)}, \Phi_{g\Delta x}^{(y)}\|\!\! =\!\!\!\int \limits_0^{2\pi}\! \frac{d\phi}{2\pi}\frac{\|\cos^2\phi , \sin^2\phi\| }{(7/6)\cos^2\phi\! +\! 0.075\sin^2\phi\! +\!g\Delta x}, \nonumber
\end{eqnarray}
where $\Phi_{g\Delta x}^{(\alpha )}$ is governed by the dimensionless parameter $g=2mT\kappa /[\sigma (\hbar\omega_f D)^2 ]$; the latter  corresponds to the linear or quadratic phonon spectra, if $g\ll 1$ or $g\gg 1$ respectively [see Eq. (17)].

Thus, the rate of momentum relaxation via phonons is proportional to the characteristic rate $\widetilde{\nu}_{ph}\propto (n_{2D}/D)^{2/3}$ while the ratios $\nu _m^{(\alpha )}/\widetilde{\nu}_{ph}\equiv F_g^{(\alpha )}$ are only dependent on the parameters of ATD through $g\propto N^{\theta -2}T/\sqrt[3]{n_{2D}^2 D}$. At $g\geq 0.1$ these dependencies are approximated as
\begin{equation}
F_g^{(\alpha )}\!\approx\!\left\{\! {\begin{array}{*{20}c}
   {1.83/(\sqrt g +0.61) ~,}\! & ~{\alpha =x}  \\
   {3.58/(\sqrt g +0.32) ~,}\! & ~{\alpha =y}  \end{array}}\! \right. ,
\end{equation}
while at $g=0$ one obtains $F_0^{(x)}\simeq$2.4 and $F_0^{(y)}\simeq$9.5. Using the above parameters $n_{2D} =10^{10}$ cm$^{-2}$, $D=$10 $\mu$m, and temperature $T=$1 K one obtains $g\sim 0.3$ for $h$-BN and $\sim 2.9$ for MoS$_2$ (the case $g<0.1$ is possible for $h$-BN devices with $D\gg$10 $\mu$m or at $T\sim$0.1 K). The temperature-independent rate $\widetilde\nu_{ph}$ is $\sim 1.4\times 10^{10}$ s$^{-1}$ for $h$-BN and $\sim 1.9\times 10^{10}$ s$^{-1}$ for MoS$_2$. For $N$-layer ATD, relaxation via flexural phonons is diminishing with $N$ because $\widetilde\nu_{ph}\propto 1/\sigma\omega_f^2\propto$ const and $\sqrt{g}\propto N^{\theta /2}$. 
\begin{figure}[t]
\begin{center}
\vspace{0.25 cm}  
\includegraphics[scale=0.27]{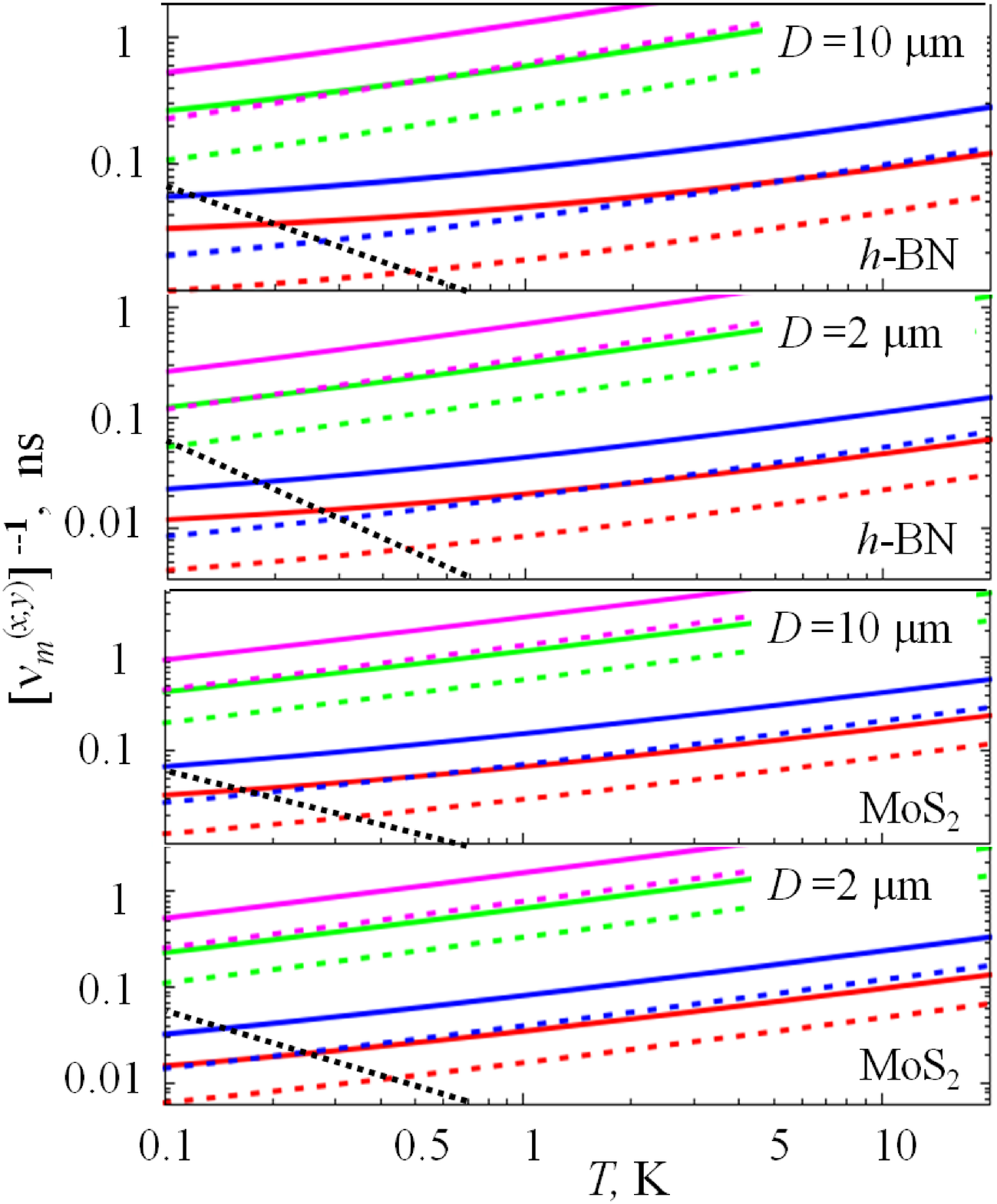}
\end{center}
\addvspace{-0.5 cm}
\caption{Phonon-induced  momentum relaxation times $[\nu_m^{(x)}]^{-1}$ and $[\nu_m^{(y)}]^{-1}$ (solid and dashed curves respectively) versus temperature for 2D-concentrations, $n_{2D}=5\times 10^8$ cm$^{-2}$ (magenta), $10^{9}$ cm$^{-2}$ (green), $5\times 10^{9}$ cm$^{-2}$ (blue), and $10^{10}$ cm$^{-2}$ (red), in $h$-BN and MoS$_2$ with $D=$10 $\mu$m and $D=$2 $\mu$m (marked). Dotted lines separate the lower left regions $\hbar\nu_m^{(\alpha )}\geq T$ where the polaron renormalization is essential. Low-concentration dependencies are cut off at temperatures corresponding to the ionization energies. }
\end{figure}

We plot the relaxation times $[\nu_m^{(\alpha )}]^{-1}$ versus $T$, which is varied over 0.1$\div$15 K, under changing of $n_{2D}$ in 20 times and at $D=$10 $\mu$m or $D=$2 $\mu$m, see Fig. 4. In the considered concentration range, $[\nu_m^{(\alpha)}]^{-1}$ changes by two orders of magnitude, since 2D-electrons become closer to ATD with increasing $n_{2D}$, but with temperature it increases by only a few times. Deviations from the $[\nu_m^{(\alpha )}]^{-1}\propto\sqrt{T}$ dependence occur at low temperatures and high concentrations moreover for the whole ($n_{2D},T$)-domain $[\nu_m^{(x)}]^{-1}>[\nu_m^{(y)}]^{-1}$ in 2$\div$3 times. Note, that the phonon-limited mobility may exceed $10^7$ cm$^2$/Vs for high temperatures and low concentrations; at $[\nu_m^{(\alpha )}]^{-1}=$1 ns the mobility is $1.8\times 10^6$ cm$^2$/Vs and the mean free pass, $\sqrt{2T/m}/\nu_m^{(ph)}$, is about 5 $\mu$m for $T=$1 K. For $N$-layer ATD these estimates approximately increase as $\sim N^{\theta /2}$ and  the mobility may exceed the results obtained for electrons on He, \cite{6} if the $rh$-scattering remains negligible. On the other hand, the broadening energy, $\hbar\nu_m^{(\alpha )}$, reaches $\sim$62 $\mu$eV or $\sim$0.7 K at the minimal relaxation times $[\nu_m^{(\alpha )}]^{-1}\sim$10 ps. Thus, the condition $\hbar\nu_m^{(\alpha )}\geq T$ is valid for low temperatures and high concentrations (in Fig. 4 this region is separated by dotted line). The polaron regime of transport appears due to the renormalization of energy spectrum similarly to the case of the 2D electrons on He interacted with the ripplon vibrations. \cite{38}

\subsection{Relaxation caused by roughness of ATD }
Next, we turn to the consideration of the roughness-induced relaxation which limits the mobility in the case of non-effective relaxation via phonons.  Leaving only contribution of the transition probability (30) to Eq. (33), one obtains the rate $\nu_m^{(rh)}$ as follows
\begin{eqnarray}
 \nu_m^{(rh)}= \frac{{2\hbar \eta _{is} \left( {f_ \bot  l_0 } \right)^2 }}{\rho _{2D} T^2 L^4}\sum\limits_{{\bf p},\Delta {\bf p}} {e^{-\varepsilon /T} \frac{{\left( {{\bf v}_{dr}  \cdot \Delta {\bf p}} \right)^2 }}{{mv_{dr}^2  }}}   \\
\times \left[ \frac{J_1 (\Delta pr_{is}/\hbar )}{\Delta p}\right]^2 \delta (\varepsilon_p  - \varepsilon_{{\bf p} - \Delta {\bf p}} )   \nonumber
\end{eqnarray}
and it does not depend on the width of trench, $D$. In analogy to the case of the phonon scattering, we perform the averaging of the $\delta$-function over the $\bf p$-plane, the averaging of $({\bf v}_{dr}\cdot\Delta{\bf p})^2$ over the $\Delta{\bf p}$-plane, and the subsequent integration over $\varepsilon$. As a result, $\nu_m^{(rh)}$ takes form
\begin{eqnarray}
 \nu_m^{(rh)}\!\! =\!\frac{\pi\eta_{is}\!\left({f_ \bot l_0}\right)^2}{{\hbar T^2 \sqrt {2m} }}\!\!\int\limits_0^\infty\!\! {d\Delta p J_1\!\!\left(\!{\frac{\Delta pr_{is}}{\hbar}}\!\right)^2\!\!\!\!\!\! \int\limits_{\Delta p^2 /8m}^\infty\!\!\!\!\!{\frac{d\varepsilon e^{-\varepsilon /T}}{\sqrt{\!\varepsilon\! -\!\!\Delta p^2\! /8m}}} }  \nonumber \\
 = \widetilde\nu_{rh}\frac{\varepsilon_{is}}{T}\!\int\limits_0^\infty\!  {d\xi}J_1\!\left(\!{\xi\sqrt {\frac{T}{\varepsilon_{is}}} }\right)^2\! e^{-\xi^2}\equiv\widetilde\nu_{rh}{\cal F}_{T/\varepsilon_{is}}  , ~~~~~ \\ 
{\cal F}_b\!\approx\! 10.8/(b+4.9)^2 , ~~ \widetilde\nu_{rh}\! =\! 2\pi^{3/2}\eta_{is}\!\left( f_\bot l_0\right)^2\! /\hbar\varepsilon_{is} ,  \nonumber
\end{eqnarray}
where we introduced the characteristic rate, $\widetilde\nu_{rh}$, the characteristic energy $\varepsilon_{is}=(\hbar /r_{is})^2 /8m$, and the dimensionless temperature-dependent function ${\cal F}_{T/\varepsilon_{is}}$. This function slowly decreas
es with the increasing of temperature, which is controlled by the characteristic energy $\varepsilon_{is}\sim $0.34 meV $\sim$4 K for $r_{is}=$50 A. For temperatures up to 15 K under consideration, ${\cal F}_b$ is decreasing from ${\cal F}_0\simeq 0.45$ to $\sim$0.2. The temperature-independent rate $\widetilde\nu_{rh}\propto n_{2D}^2\eta_{is}$ is $\sim 0.44\times 10^{10}$ s$^{-1}$ for $h$-BN and $\sim 1.6\times 10^{10}$ s$^{-1}$ for MoS$_2$ at the above-used parameters ($n_{2D}\sim 10^{10}$ cm$^{-2}$, $\eta_{is}\sim 10^{-3}$ , and $r_{is}=$50 A). 
 
In Fig. 5 we plot the relaxation times $[\nu_m^{(rh)}]^{-1}$ for the ($n_{2D},T$)-domain under consideration at $r_{is}=$50 A or 100 A; since $\nu_m^{(rh)}\propto\eta_{is}$ we choose $\eta_{is}=10^{-3}$ when an interplay between $ph$- and $rh$-induced relaxations is essential. In analogy to the scattering via phonons, $[\nu_m^{(rh)}]^{-1}$ changes about two orders of magnitude under variation of $n_{2D}$ but the relaxation times for $r_{is}=$50 A are temperature-independent at $T<$1 K while at $T>$5 K $[\nu_m^{(rh)}]^{-1}$ increases approaching to $T^2$-dependency. With increasing $r_{is}$ to 100 A, the relaxation times decrease in a several times at $T\simeq$0.1 K but at $T\geq$10 K they are the same order due to an increasing with $T$. In contrast to the case of the $ph$-scattering, now $[\nu_m^{(rh)}]^{-1}\geq$0.1 ns for the conditions considered and the renormalization of mass is only possible for ATD with a strong roughness, if $\eta_{is}>10^{-2}$. 

For the case of monolayer ATD with the parameters of roughness used, conditions of interplay between $ph$- and $rh$-channels of relaxation are clear from comparison of Figs. 4 and 5.  For $N$-layer ATD or different parameters of roughness, one should re-scale Figs. 4 or 5 taking into account that $[\nu_m^{(\alpha )}]^{-1}\propto N^{\theta /2}$ or $[\nu_m^{(rh)}]^{-1}\propto 1/\eta_{is}$  respectively. In the case of an arbitrary shape islands, more complicate ${\cal W}_{{\bf p},{\bf p}'}^{(rh)}$ should be used in Eq. (33). But the rate $\nu_m^{(rh)}$ is expressed through the  area of roughness and the characteristic size of islands which are similar to the parameter $\eta_{is}$ and the function ${\cal F}_b$. 
\begin{figure}[t]
\begin{center}
\vspace{0.25 cm}  
\includegraphics[scale=0.24]{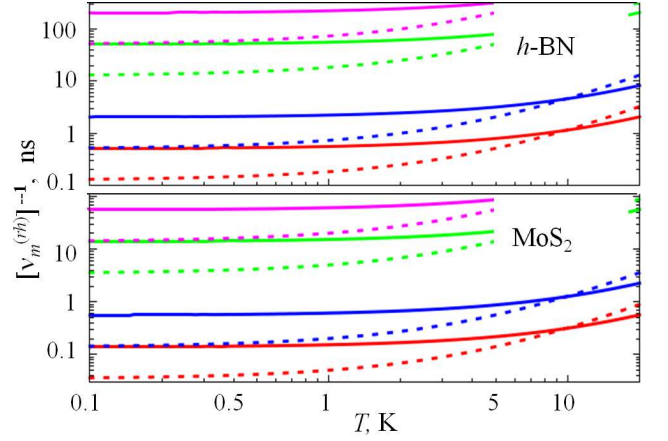}
\end{center}
\addvspace{-0.5 cm}
\caption{Roughness-induced  momentum relaxation times $[\nu_m^{(rh)}]^{-1}$ versus temperature for 2D-concentrations, $n_{2D}=5\times 10^8$ cm$^{-2}$ (magenta), $10^{9}$ cm$^{-2}$ (green), $5\times 10^{9}$ cm$^{-2}$ (blue), and $10^{10}$ cm$^{-2}$ (red) in $h$-BN and MoS$_2$ with disk islands of radius $r_{is}=$50 A and 100 A (solid and dashed curves respectively) at $\eta_{is}=10^{-3}$; note that $[\nu_m^{(rh)}]^{-1}\propto 1/\eta_{is}$.  Similarly to Fig. 4, the low-concentration dependencies are cut off at $T\sim{\cal E}_i$. }
\end{figure}

\subsection{Energy relaxation }
Here we turn to consideration of the losses of energy, $Q$, determined by Eq. (32b). After the expansion of $ Q $ in small temperature change for the weak heating case, $T\gg |T_e -T|$, the energy relaxation rate takes form
\begin{equation}
\nu_e =\sum\limits_{{\bf p},{\bf p}'}\frac{(\varepsilon  - \varepsilon ')^2}{n_{2D} L^2T^2}\Delta {\cal W}_{{\bf p},{\bf p}'}\widetilde f_\varepsilon ~,
\end{equation}
moreover $\Delta {\cal W}_{{\bf p},{\bf p}'}^{(ph)}$ is given by Eq. (27b) while elastic processes drop out from $\nu_e$. The straightforward transformations of Eq. (39) yield the double integral for this rate
\begin{equation}
\nu_e\! =\!\frac{\pi\hbar f_\bot^2 \rho_{2D} }{4\sigma T^2}\!\int\limits_0^\infty\! {d\varepsilon }\!\int\limits_0^\infty\!{d\varepsilon '} e^{-\varepsilon /T} (\varepsilon -\varepsilon ')^2\delta ''(\varepsilon -\varepsilon ') , 
\end{equation}
where the $\propto\delta '(\dots )$ term of Eq. (27b) gives zero contribution to $\nu_e$. The result of integration is $\nu_e =f_\bot^2 m/(2\hbar\sigma T)$ and the energy relaxation time $\nu_e^{-1}\propto NT/n_{2D}^2$. Since $\nu_e /\widetilde{\nu}_{ph}=2m(2\omega_f D)^2 /\sqrt{\pi}T\ll 1$ and this ratio $\propto 1/NT$, the energy relaxation time appears to be $\sim 2\div 4$ orders longer in comparison to the momentum one. It means that $\nu_e^{-1}$ may increase up to a microsecond time range for high $T$ and low $n_{2D}$ or in a multilayer ATD.

A simple way to examine of $\nu_e$ is the Joule heating of 2D electrons with an increasing of the in-plane electric field $E_\alpha$ ($\alpha =x$ or $y$ for nonlinear transport along or across trench) which can be described by the energy balance per electron. Here we do not study the current-voltage characteristic of the device but only discuss a condition for the linear regime, $T_e -T\ll T$. Equating the increase of energy per electron due to the Joule heating, $(eE_\alpha )^2 /m\nu_m^{(\alpha )}$, and its losses, $Q$, we find the temperature change $T_e -T=(eE_\alpha )^2 /m\nu_m^{(\alpha )}\nu_e$. The linear response takes place under fields restricted by the condition
\begin{equation}
E_{\alpha}\ll\sqrt {Tm\nu_e\widetilde{\nu}_{ph}F_g^{(\alpha )}}/|e|\equiv\widetilde{E}_{\alpha} .
\end{equation}
The limiting field $\widetilde{E}_{\alpha}$ is dependent on $T$ through $\sqrt{F_g^{(\alpha )}}$ determined by Eq. (36) while $\sqrt{Tm\nu_e\widetilde{\nu}_{ph}}\propto (n_{2D}/D)^{4/3}$ does not depend on $T$. For a few layer ATD, $\widetilde{E}_{\alpha}$ decreases with increasing of $N$ 
because $\nu_e\propto 1/N$ and $g\propto N^{\theta /2}$.

The temperature and concentration dependencies of the limiting field $\widetilde{E}_{\alpha}$ are plotted in Fig. 6 for the monolayer ATDs with different $D$.  Similarly to Fig. 4, this field varies in a few times with $T$ and changes with $n_{2D}$ in about two orders. At low $n_{2D}$ and $T>$1 K $\widetilde{E}_\alpha$ is dropped up to $\leq$ mV/cm so that the linear regime is restricted by the voltages $\sim\mu $V applied to a device of lenght $\sim 100$ $\mu$m; in addition $\widetilde{E}_{\alpha}$ decreases at $N>1$. A diminution of $\widetilde{E}_{\alpha}$ is also restricted due to the contribution of the $rf$-induced momentum relaxation, when the complete rate $\left[\nu_m^{(\alpha )}+\nu_m^{(rf)}\right]$ determines the Joule heating, or due to the ionization processes, if $T_e$ is comparable to ${\cal E}_i$.
\begin{figure}[t]
\vspace{0.27 cm}
\begin{center}
\includegraphics[scale=0.27]{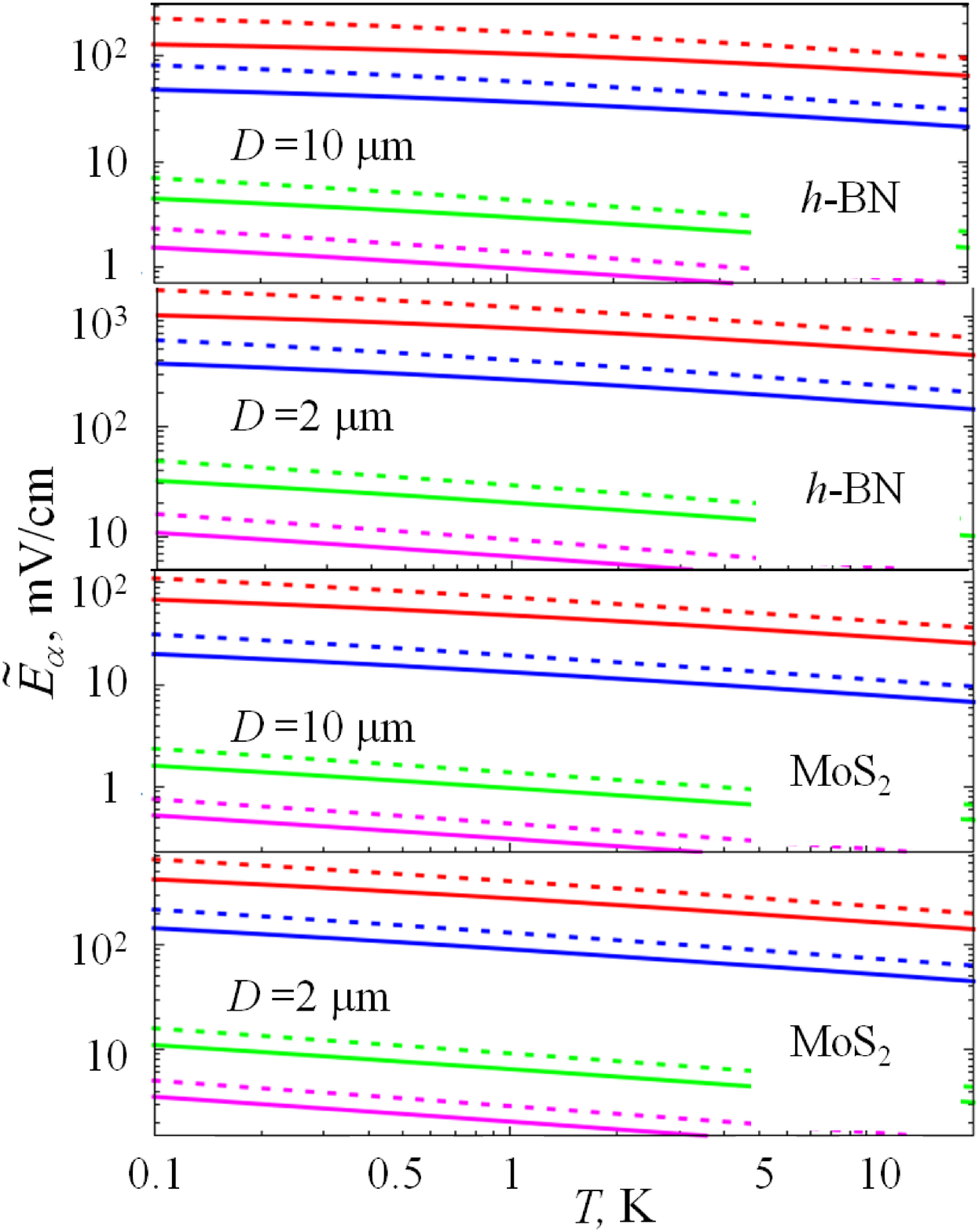}
\end{center}
\addvspace{-0.5 cm}
\caption{Temperature dependency of fields $\widetilde{E}_x$ and $\widetilde{E}_y$ (solid and dashed curves respectively) for 2D concentrations $n_{2D}=5\times 10^8$ cm$^{-2}$ (magenta), $10^{9}$ cm$^{-2}$ (green), $5\times 10^{9}$ cm$^{-2}$ (blue), and $10^{10}$ cm$^{-2}$ (red), in $h$-BN and MoS$_2$ with $D=$10 $\mu$m and $D=$2 $\mu$m (marked). Low-concentration dependencies are cut off at temperatures corresponding to the ionization energies. }
\end{figure}

\section{Concluding remarks }
Summarizing of the consideration presented, the examination  of the 2D  electrons trapped in vacuum near the ATD suspended above the back gate is performed here. It is found that the Boltzmann liquid of 2D electrons  floated on ATD arises at temperatures 0.1$\div$15 K under the bias fields 0.8$\div$18 kV/cm which correspond the concentrations $n_{2D}=5\times 10^8\div 10^{10}$ cm$^{-2}$. The leakage current through a perfect ATD is weak and the polarizability induced by electrons in a few-layer ATD  is negligible. The long-wavelength crossover from the quadratic dispersion law of the flexural vibrations to the linear one appears due to the bending of ATD under pressure of 2D electron  caused by attraction from the back gate. The in-plane transport is limited both these flexural phonons and the monolayer islands randomly placed on ATD. The momentum and energy relaxation rates vary in about two orders over the interval of $n_{2D}$ considered but in only several times with temperature (notice, that Figs. 4-6 are plotted in the double-logarithmic scale). For the low $T$ and high $n_{2D}$, the polaron renormalization of mass is essential, i.e. one deal with the Boltzmann liquid of polarons. Contrary, at high $T$ and low $n_{2D}$ the  phonon scattering is suppressed and the mobility reaches a 10-million range, if the $rh$-scattering remains weak. For $N$-layer ATD the relaxation via flexural phonons is suppressed and the scattering via roughness becomes dominant. The quasielastic relaxation of energy reachs up to a microsecond time scales and the region of linear response is restricted by the in-plane electric fields 1$\div 100$ mV/cm.

The study is based on a several assumptions which are listed and shortly discussed below. {\bf (a)} A rough estimate of the tunnel leakage rate in Sec. IIA justifies the zero boundary condition for Eq. (18a) and shows that the implementation and verification of the 2D electrons on ATD is possible. A reliable study of this process for electrons with energies $<0.1$ eV requires a direct measurement of the leakage current and an exact microscopic calculation. Effects caused by an imperfections of ATD, such as leakage of 2D electrons through microscopic holes or their localization at capture centers are not considered here. {\bf (b)} Neglecting of ATD polarization in Eq. (18b) is based on the estimate of Sec. IIB for  $N\leq 5$. The question about a thickness of ATD when the image force becomes essential remains open and a more careful study is of interest. {\bf (c)} Self-consistent description of the energy spectrum gives good estimate of the ground-state and ionization energies but a more precise calculations of the excited levels are necessary for study of the microwave response. {\bf (d)} The phonon spectrum is analyzed for the case of the ATD with clamped edges but a heat exchange through the edges is not considered.  Supposing that this exchange is strong enough we apply the equilibrium phonon distribution. {\bf (e)} The study of in-plane transport, which is based on the balance equations for momentum and energy, gives an approximate estimation of the relaxation times. Nevertheless due to the strong dependencies on $T$ and $n_{2D}$, these results open a way for characterization of the scattering mechanisms (with an adding of other channels of relaxation, e.g. charged imperfections in ATD or noise from the back gate). {\bf (f)} Peculiarities of the charge transfer through contacts as well as the boundary conditions at the side edges of ATD suspended over trench were not considered but these factors may be essential for small-size devices. {\bf (g)} Beyond the ($T, n_{2D}$)-region considered, the analysis should be more complicated. In principle, theories for the Boltzmann 2D liquid of polarons or for the ballistic transport of this liquid, which should be based on the nonequilibrium diagram technique, are timely but more information on a parameters of device is necessary. To finish this list stress that all the above-discussed assumptions and restrictions do not change the results and conclusions of the analysis performed. 

Next, there are some comments on a possibility for realization of the device suggested. It seems, that it is not a difficult technological problem to produce the ATD suspended over the back gate  and merged to the lateral contacts for 2D electrons. \cite{16,39,40} The control and characterization of such a device should be similar the case of 2D electrons on liquid He. \cite{41} Differences in parameters of the $h$-BN- and MoS$_2$-based structures demonstrate that improvement of their characteristics by using different ATDs is possible and such a way for optimization of the device would be useful. The upper temperature restriction due to the low energy of ionization can be avoided by implementing an additional top gate above ATD, which provides a more tunable discrete energy spectrum. Beside of this, one can consider an implementation of a  double-ATD structure separated on a hundred(s) A, when 2D electrons are confined between these ATDs. An inhomogeneous back (or top) gate permits one to modulate of 2D concentration, including a realization of the 1D electrons or the lateral array of quantum dots. These trapped electrons can serve as the qubits of a quantum computer, see analysis \cite{8} for electrons trapped over liquid He. 

To conclude, an implementation of 2D electrons confined in vacuum over the ATD seems to be quite possible technologically. A study of the arising Boltzmann plasma should demonstrate new physical characteristics which vary greatly with temperature and concentration. There is a potential for an application in modern (opto)electronics both for the simple device analyzed and for a more complicate structures mentioned above. When implementing the non-uniform gate(s), a possibility is opened for the new type of quantum hardware using a qubit which is based on the single electron. Because of trapping in vacuum over the ATD such a qubit  is isolated better from an environmental noise.   \\

The data that supports the findings of this study are available within the article and from the author upon request.



\begin{thebibliography}{52}
\bibitem{35}
Under the variational calculation, the energy of the next level, ${\cal E}_1$, turns out to be close to the ionization level; if we use the linear potential $W_z\propto z$ at $z>0$, ${\cal E}_1$ exceeds ${\cal E}_i$. Hence without the upper gate, it is enough to consider the ground level and to determined the energy of ionization of the 2D electrons into the continuous spectrum.
\bibitem{41}
Because 2D-electrons are trapped due to attraction by the back gate, a question arises on heat removal from an initially formed capacitor which is uncharged (possible extra charge is moved out of device). There are two possible mechanisms: a/ heat transfer due to the relaxation of hot electrons via flexural phonons of ATD and b/ electron-electron scattering with an escape of high-energy electron out of device and a capture of low-energy one (similarly to the Auger process). A relaxation time via the channel a/ may exceed the results for 2D-electrons (Sec. VC) in 1$\div$2 orders but it is much shorter than the leakage times, $\Gamma^{-1}$ (Sec. IIA), so that there is no a bottleneck for trapping of 2D-electrons. An additional channel b/ may be dependent on different factors (initial energy and concentration of electrons, geometry of device, etc.) and this mechanism should be analyzed separately.
\end{thebibliography}
\end{document}